\documentclass[12pt,preprint]{aastex}

\def\ltsima{$\; \buildrel < \over \sim \;$}
\def\simlt{\lower.5ex\hbox{\ltsima}}            
\def\gtsima{$\; \buildrel > \over \sim \;$}
\def\simgt{\lower.5ex\hbox{\gtsima}}            

\newcommand{\asca}{{\it ASCA}}
\newcommand{\rosat}{{\it ROSAT}}

\newcommand{\heao}{{\it HEAO1}}
\newcommand{\chandra}{{\it Chandra}}
\newcommand{\xmm}{{\it XMM-Newton}}

\newcommand{\ergs}{erg cm$^{-2}$ s$^{-1}$}

\newcommand{\nh}{$N_{\rm H}$}

\slugcomment{July 22}
\shorttitle{Optical Identification of the {\it ASCA} Lynx Survey}
\shortauthors{Ohta et al.}

\begin{document}

\title{Optical Identification of the {\it ASCA}
Lynx Deep Survey: An Association of QSOs and a Supercluster
at $z=1.3$?\footnote{
Based on observations made with the Kiso Observatory, which is
 operated by Institute of Astronomy, the University of Tokyo, 
with the Kitt Peak National Observatory 2.1m and 4m telescopes,
 which are operated by National Optical Astronomy Observatories (NOAO)
 operated by AURA, Inc., under contract with the 
National Science Foundation, with the University of Hawaii 88'' telescope,
and with the William Herschel Telescope  operated
 on the island of La Palma by the Isaac Newton Group in the Spanish 
Observatorio del Roque de los Muchachos of the Instituto de 
Astrofisica de Canarias. 
}
}

\author{Kouji Ohta,\altaffilmark{1,2,3} 
Masayuki Akiyama,\altaffilmark{2,3,4}
Yoshihiro Ueda,\altaffilmark{5}
Toru Yamada,\altaffilmark{2,3,6}
Kouichiro Nakanishi,\altaffilmark{2,7}
Gavin B. Dalton,\altaffilmark{8,9}
Yashushi Ogasaka,\altaffilmark{10}
Tsuneo Kii,\altaffilmark{5}
and
 Kiyoshi Hayashida\altaffilmark{11}
}

\altaffiltext{1}{Department of Astronomy,
Kyoto University, Kyoto 606-8502, Japan.}
\altaffiltext{2}{Visiting Astronomer, Kitt Peak National
Observatory, National Optical Astronomy Observatories (NOAO).
NOAO is operated by AURA, Inc., under contract with the 
National Science Foundation.}
\altaffiltext{3}{Visiting Astronomer, University of Hawaii Observatory.}
\altaffiltext{4}{Subaru Telescope, National Astronomical Observatory
of Japan, Hilo, HI, 96720.}
\altaffiltext{5}{Institute of Space and Astronautical Science, 
Kanagawa 229-8510, Japan.}
\altaffiltext{6}{National Astronomical Observatory of Japan, Tokyo
 181-8588, Japan.}
\altaffiltext{7}{Nobeyama Radio Observatory, National Astronomical
 Observatory of Japan, Nagano 384-1305, Japan.}
\altaffiltext{8}{Department of Astrophysics, University of Oxford,
 Oxford OX1 3RH, UK.}
\altaffiltext{9}{Rutherford Appleton Laboratory, Chilton, Didcot
OX11 0QX, UK}
\altaffiltext{10}{Department of Physics, Nagoya University, Nagoya
 464-8602, Japan.}
\altaffiltext{11}{Department of Earth and Space Science, Osaka
University, Osaka, 560-0043, Japan.}

\begin{abstract}
Results of optical identification of the \asca\  Lynx deep survey are
presented.
Six X-ray sources are detected in the  2--7 keV band using the SIS 
in a $\sim 20^{\prime} \times 20^{\prime}$ field of view
with fluxes larger than $\sim 4 \times 10^{-14}$
 erg s$^{-1}$ cm$^{-2}$ in the band.
Follow-up optical spectroscopic observations were made, and
five out of six sources are identified with AGNs/QSOs at
redshifts of 0.5 -- 1.3.
We also identify two more additional X-ray sources detected in
a soft X-ray band with AGNs/QSOs.
It is found that  three QSOs identified are located at $z\sim 1.3$.
Two rich clusters and several groups of galaxies are also placed
at the same redshift 
in the surveyed field, and projected separations between the QSOs
 and the clusters are 3--8 Mpc at the redshift.
\end{abstract}

\section{INTRODUCTION}

The optical identification of X-ray sources is an important step
to  reveal cosmological evolution of AGN/QSOs as well as
the evolution of super-massive black holes harbouring in their 
host galaxies.
It is also a direct way to solve the origin of cosmic
X-ray background (CXB).
Although many optical identifications of soft X-ray selected
samples have been already made (e.g., McHardy et al. 1998;
Lehmann et al. 2001), 
it is essential to use a hard X-ray selected sample, because
the energy of the CXB peaks at around 20 keV and the bulk of
the CXB energy is thought  to come from moderately
obscured ($N_{\rm H} \sim 10^{23-24}$ cm$^{-2}$) type-2 AGNs
(e.g., Comastri et al. 1995; Ueda et al. 2003). 
The soft X-ray emission such as in the 0.2 -- 2 keV band,
which was used in the \rosat\  surveys, 
is insensitive to sources with such high column densities,
which can only be detected in the harder X-ray bands.
Therefore the optical identification of X-ray sources
selected in a hard X-ray band (e.g., 2--10 keV) is required  
to examine the evolution of type-1 and -2 AGNs and to solve the
 origin of the CXB.

In this context, we have been pursuing optical follow-up
spectroscopy of hard X-ray selected sources.
Akiyama et al. (2000) obtained optical identifications
for the X-ray sources detected by \asca\  with a flux
larger than $\sim 1 \times 10^{-13}$ erg s$^{-1}$ cm$^{-2}$
in a 2-10 keV band (Ueda et al. 1999). 
The survey area was a contiguous sky area ($\sim 5$ deg$^2$) 
near the north Galactic pole, and we identified 33 X-ray sources
among 34 sources.
We have obtained further optical identifications
of a subsample of the \asca\  Medium Sensitivity Survey
(Ueda et al. 2001).
The survey is a serendipitous source survey making use of
many \asca\  pointing observations made during a three year
period; from this catalog we selected 87 hard X-ray selected
sources and have identified all of them (Akiyama et al. 2003).
Although these surveys are shallower by about two orders of
magnitude than  those made by \chandra\  and \xmm,
the resultant catalogs  cover the flux level between
these very deep surveys and the much shallower survey
previously made by \heao\  ($\sim 10^{-11}$ erg s$^{-1}$ cm$^{-2}$
in the 2--10 keV band), and our catalogs play a unique role to
investigate the cosmological evolution of AGN/QSOs; the evolution
of the luminosity function in hard X-ray band, the luminosity
dependence of the $N_{\rm H}$ distribution, and the 
origin of the CXB
(e.g., La Franca et al. 2002; Cowie et al. 2003; Ueda et al. 2003).

In this paper, we report results of our  optical identification
for the \asca\ Lynx deep survey.
Our survey  covers a wider field than the \chandra\ survey of 
Stern et al. (2002) made in 2000 May.
The optical identification of the Stern et al.  Lynx survey
remains incomplete, with only a relatively small
number of sources having been identified.
We therefore believe that adding further optical
identifications for hard X-ray band selected AGN/QSOs with a 
flux level of $10^{-13} \sim 10^{-14}$ erg s$^{-1}$ cm$^{-2}$
(2--10 keV) would be still valuable and useful, though our
sample size is very small.
Preliminary results were presented by Ohta et al. (1998),
and a particularly interesting result concerning a
discovery of a type-2 QSO candidate was described by Ohta et al. (1996)
(with subsequent results by Akiyama, Ueda, \& Ohta (2002)).

\section{X-RAY DATA}
The \asca\ deep observation of the Lynx field was performed from
1993 May 13 12:55 (UT) to May 15 19:13. We here concentrate on the
data of the SIS instrument (Burke et al. 1991), which was operated in
the 4-CCD Faint Mode. The total field of view (FOV) covered by the
SIS0 and SIS1 had roughly a square shape with corners corresponding to
(RA, DEC(J2000)) = ($8^{\rm h} 50^{\rm m} 20.^{\rm s}4$, $+44^{\circ}
59^{\prime} 27^{\prime\prime}$), ($8^{\rm h} 48^{\rm m} 6.^{\rm s}6$,
$+45^{\circ} 2^{\prime} 43^{\prime\prime}$), ($8^{\rm h} 47^{\rm m}
49.^{\rm s}4$, $+44^{\circ} 40^{\prime} 7^{\prime\prime}$), and
($8^{\rm h} 50^{\rm m} 2.^{\rm s}4$, $+44^{\circ} 36^{\prime}
52^{\prime\prime}$). A total survey area was about 530 arcmin$^2$. An
average net exposure of 82 ksec (per one CCD) was obtained after
standard data screening. We made sky images in the 3 energy bands,
0.7--7 keV, 0.7--2 keV, and 2--7 keV, from events of grade 0, 2, 3,
and 4, discarding the very edge regions of each chip to avoid complexity
due to satellite-attitude fluctuations. Note that the sensitivity was
not uniform over the FOV also because of vignetting and gaps between
the CCD chips. We calibrated the absolute fluxes by a correction
factor of 10\% so that they match with those obtained by the GIS
(Ohashi et al. 1996), referring to the results of the CXB spectrum
analysis from the same data by Miyaji et al. (1997).

To produce an X-ray source list, we adopted the same procedure as
applied to the \asca\ deep survey data in the Lockman hole field
(Ishisaki et al. 2001), which utilized the positions of \rosat\
sources. In this method we did not perform source detection initially.
Instead, we determined fluxes at positions of known (soft) X-ray
sources, utilizing the image fitting method where the detector
response was taken into account (Ueda et al. 1999). This technique is
effective to minimize source confusion when we have sufficiently deep
(and wide FOV) \rosat\ data. Analysis was made in the three energy
bands. For the initial source list we adopted the WGA catalog of PSPC
sources (White, Giommi, \& Angelini 1994) available from the HEASARC
database. We then repeated the fitting process by adding new sources
to the input list when any significant peaks were found in the
residual image. This procedure was iterated until we confirmed that
there was no significant source above 3.5 $\sigma$.

Table~1 gives the complete X-ray source list from the \asca\
observation, containing sources detected with significance above
3.5$\sigma$  in any of the three bands, sorted by RA and DEC. The count
rate has been converted into the energy flux of the same band
by assuming a power law photon index of 1.6.
There are two sources detected with \asca\ but not with \rosat.
The positional errors of these \asca\ only
sources are conservatively estimated at $0.'5$.

From this list, we defined a hard-band selected sample consisting of 6
sources significantly (larger than 3.5$\sigma$) detected in the 
2--7 keV band, which are marked with * in Table~1.
Using the \chandra\ archival data, we made spectral analysis
of the three sources (No.\ 1, No.\ 9, and No.\ 11) from this
sample which are  located within the \chandra\ FOV. 
We corrected the level-1 event data for the Charge Transfer
Inefficiency using the technique developed by Townsley et al. (2000)
and used the appropriate energy response.
 Spectral fits were  performed using the redshift information
 from optical identification described below.
The type-2 quasar candidate AX~J08494+4454 (No.\ 11)
shows a large absorption of \nh\ $\simeq 2\times10^{23}$ cm$^{-2}$
(Akiyama et al. 2002), while the other two (No.\ 1 and No.\ 9)
show a power law with almost no absorption over that expected
from the  Galaxy. 
The results are summarized in Table~2.

\section{OPTICAL IDENTIFICATION}

In order to identify candidate optical counterparts of the 
{\it ASCA} sources, 
we conducted optical imaging observations with Kiso 1.05m Schmidt
telescope with a Tek $1024\times1024$ CCD on January 1995 and 
follow-up deeper $I_{c}$ band imaging observations 
with a Tek $2048\times2048$ CCD attached to the University of Hawaii
88$^{\prime\prime}$ telescope  on March 1995 and April 1996.
In the latter observations, one CCD pixel corredponded to
0.$^{\prime\prime}$22 and a seeing (FWHM)
during the observing runs was  typically 1.$^{\prime\prime}$5.
Exposure times were between  5 and 20 minutes.
Landolt's standard stars (Landolt 1992) were also observed a few times
in each night.

\begin{figure}
\plotone{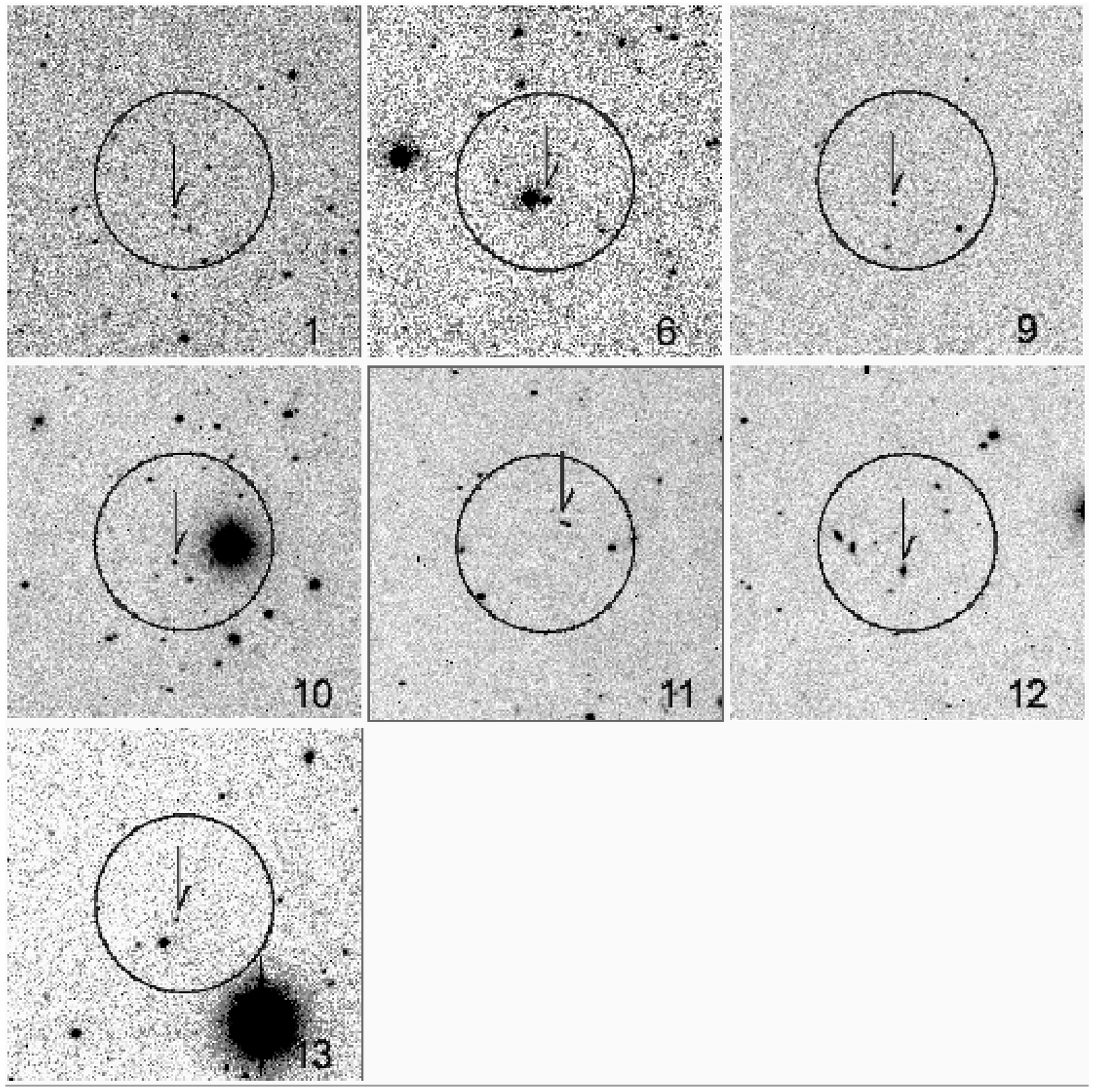}
\caption{
$2'\times 2'$ optical ($I_{\rm C}$ band) images of 
\asca\ hard-band selected sources in the Lynx field
taken with the UH88$^"$ telescope.
North is to the top and East is to the left.
Large circles indicate the $0.'5$ error circles of the X-ray
sources.
An arrow marks the optical counterpart of each X-ray source.
Note that No.\ 6 and No.\ 13 are not included in the hard X-ray
selected sample; they are additional objects detected in
the soft band.
\label{fa1}}
\end{figure}

The $2^{\prime}\times2^{\prime}$ $I_{c}$-band images for the 
hard-band selected sources are shown in Figure~\ref{fa1},
with the exception of source No.\ 2, 
for which we  do not have imaging data.
We also show images for No.\ 6 and No.\ 13, which are not hard
X-ray selected objects, but which were identified in the following
spectroscopic observations.
The large circle in each image indicates the $0.'5$ error circle
of the X-ray source.

We conducted  optical spectroscopy of  the optical
candidates  with the Gold Camera
Spectrograph on the Kitt Peak National Observatory (KPNO) 
2.1m telescope in February 1995.
We used a grating of 158 mm$^{-1}$ blazed at 6750{\AA} and
a WG345 filter for the order cut. 
A wavelength range from 5000{\AA} to 10000{\AA} was covered
with a spectral resolution of 13{\AA}.
The slit width  was 2$^{\prime\prime}$.
One CCD pixel corresponded to 2.5 {\AA} for the dispersion
direction and 0.$^{\prime\prime}$78 for the spatial direction.
The typical seeing during the observations was $\sim
 2.^{\prime\prime}5$.
We also conducted  optical spectroscopic observations with the
 multislit spectrograph (CryoCam) on the KPNO Mayall 4m
 telescope in April 1996. 
We used the grism 730 to covere a spectral range
from 6000 {\AA} to 9000 {\AA} with a spectral resolution of 22 {\AA}.
The slit width used was 2.$^{\prime\prime}5$.
One CCD pixel corresponded to 4.3 {\AA} and 0.$^{\prime\prime}84$.
The seeing was typically 1$^{\prime\prime}-2^{\prime\prime}$.

We took spectra of candidates of optical counterparts
of all the hard-band selected sources, except for No.\ 2.
We also observed two additional sources (No.\ 6 and No.\ 13),
though they are not hard band selected objects
(their significance levels in the hard band are 3.0 and 3.1.)
An exposure time of each frame in these observations was 20--30
minutes.
A total exposure time for each target ranged from 30 to 90 minutes.
In all spectroscopic observations including those described below,
we took spectra of KPNO spectrophotometric standard stars.
The data were reduced with the usual manner using IRAF;
after debiasing, the flat fielding was carried out and the
wavelength calibration was made using HeNeAr lamp data.
The background was subtracted with BACKGROUND  task and
a sensitivity correction was applied using spectra of the 
spectrophotometric standard stars.
Finally, the spectrum of the target object was extracted using
APALL task by tracing the spectral continuum.

In these observations,  we found an object which showed AGN-like
emission lines for each of the X-ray sources.
Some of them showed only one broad emission line.
They are marked with arrows in Figure~\ref{fa1}, and their
coordinates and optical magnitudes are listed in Table~2.
Astrometry was done using APM catalog (Irwin, Maddox, \&
McMahon 1994).
The $I_{\rm C}$-band magnitude was derived using Landolt's
standard stars and the photometric error is estimated to
be 0.03 mag to 0.2 mag.
Magnitudes of three of them are given by Stern et al. (2002)
and agree with our magnitudes within the errors.

In order to obtain higher SN spectra and achieve secure 
redshift determination with additional emission lines,
we obtained further spectroscopic observations with the Wide Field 
Grism Spectrograph (WFGS) on the UH88$^{\prime\prime}$ in 
March 1998, and the ISIS spectrograph on the 4.2m William 
Herschel Telescope (WHT) in December 1998 (service observation).
No.\ 1, No.\ 9, No.\ 10,  and No.\ 12 were observed with the WFGS 
with a grating of 420 mm$^{-1}$ blazed at 6400{\AA}.
A slit width of 1.$^{\prime\prime}2$ was used.
The spatial sampling was 0.$^{\prime\prime}35$ pixel$^{-1}$, and
the seeing size was typically 1$^{\prime\prime}$.
The wavelength range from 4000{\AA} to 9000{\AA} was covered 
with a spectral resolution of 12{\AA}. 
No.\ 10 and No.\ 13 were observed with the ISIS two beam
spectrograph with R300B and R158R gratings on the blue and red arms, 
respectively.
In combination with the two arms, a wavelength range from 
3300{\AA} to 8000{\AA} was covered with spectral resolutions 
of 6{\AA} (blue) and 10{\AA} (red).
A total exposure time was 20 minutes for each of the target.
The data were reduced in the same way as described above.
In order to improve the signal-to-noise ratios
 for No.\ 1 and No.\ 12, we averaged two
one-dimensional spectra obtained in different observing runs
(UH88$^{\prime\prime}$ and KPNO 4m for No.\ 1 and UH88$^{\prime\prime}$ and
KPNO 2.1m for No.\ 12) after adjusting the differences of spectral resolutions
by convolving Gaussian kernels.

\begin{figure}
\epsscale{0.6}
\plotone{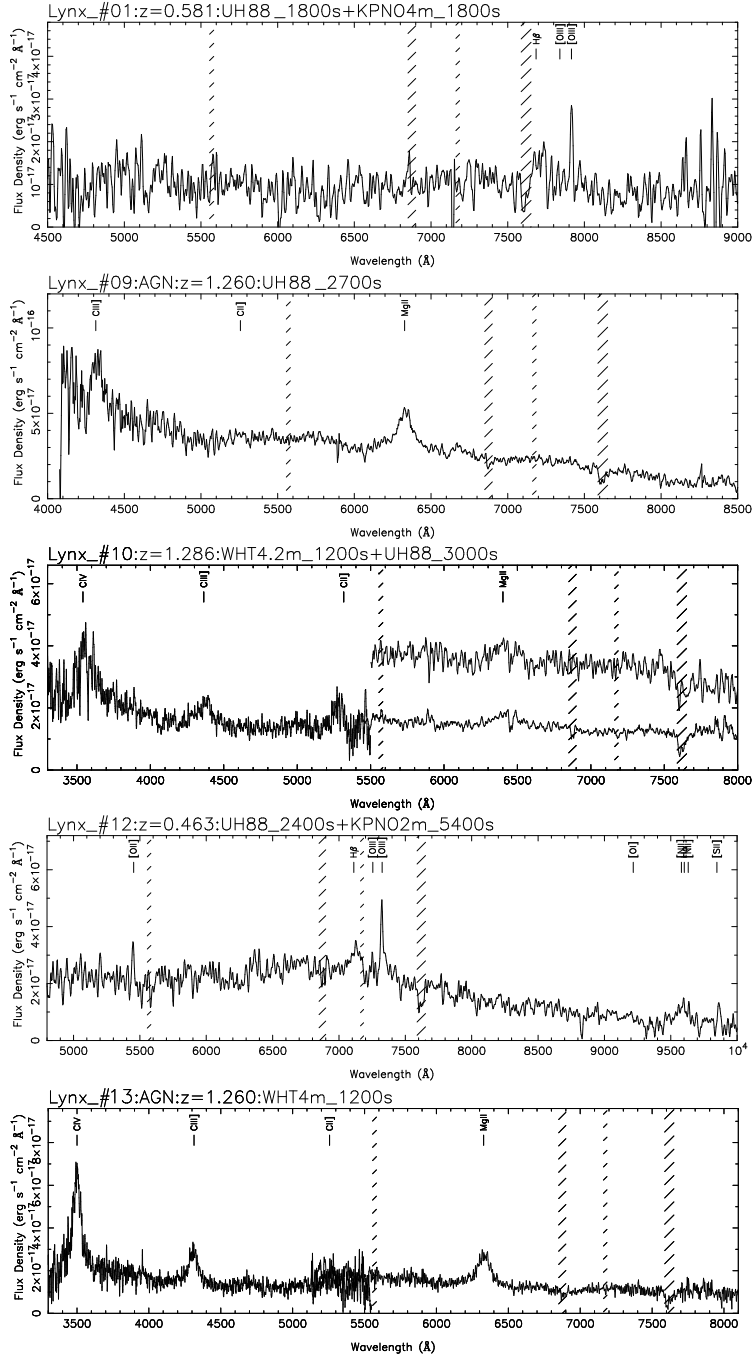}
\caption{
Optical spectra of the  counterparts of
\asca\ hard-band selected sources and of No.\ 13.
Identifications of the detected emission lines are marked with
vertical lines. Hatched areas represent the
wavelength ranges affected by strong night sky lines or 
atmospheric absorptions.
For No.\ 10, a spectrum taken with the WHT is plotted 
and that taken with the UH88$^"$ is plotted in the
upper part (offsetted) to show the absorption feature
 in the MgII emission line independently.
\label{fa2}
}
\end{figure}

The resultant optical spectra of the counterparts of
the hard X-ray selected objects are shown in Figure~\ref{fa2}.
Object No.\ 1 shows an H$\beta$ emission line and an 
[OIII]$\lambda$5007 emission line.
Although the signal-to-noise ratio is not so good,
we measured a line width of the H$\beta$ emission line
to be $4000 \sim 5000$ km s$^{-1}$, thus identified it
with a broad emission line.
The redshift obtained is 0.581.
No.\ 9 shows a  broad (9200 km s$^{-1}$) MgII$\lambda$2800 
emission line  and a CIII]$\lambda$1909 emission line with a 
redshift of 1.260.
No.\ 10 shows  broad (14000 km s$^{-1}$) MgII$\lambda$2800,
CII]$\lambda$2326, CIII]$\lambda$1909, 
and CIV$\lambda$1549 emission lines.
The redshift is 1.286. 
In the MgII$\lambda$2800 emission line, there seems 
to be an absorption feature.
No.\ 11 has a very hard X-ray spectrum and is identified with 
a type-1.9 QSO at $z=0.886$, showing narrow
H$\beta$,  strong [OIII]$\lambda\lambda$4959/5007, and
[NeV]$\lambda$3426 as well as some other [NeIII] lines (Ohta
et al.1996), but showing a broad H$\alpha$ emission (Akiyama et al.
2002).
We do not reproduce the spectrum of this object in this paper.
No.\ 12 shows a broad H$\beta$ emission line ($\sim 3400$ km s$^{-1}$)
and a strong [OIII]$\lambda$5007 and is identified 
with an AGN at $z=0.463$.
[OII]$\lambda$3727 is also seen clearly.
The results are summarized in Table~2.
In Table~2, an X-ray luminosity refers to an intrinsic luminosity
in the rest-frame 2--10 keV band after correcting absorption
by using an SIS 2--7 keV count rate and the photon indices
listed in Table 2.
We adopt a cosmological parameter set of $H_0 = 70$ km s$^{-1}$ 
Mpc$^{-1}$, $\Omega_{\rm M} =0.3$, and $\Omega_{\Lambda} = 0.7$.
These X-ray luminosities are used by Ueda et al. (2003).

We also took optical spectra of No.\ 6 and No.\ 13 with
KPNO 2.1m and WHT, respectively.
No.\ 6 shows  broad H$\beta$ and strong 
[OIII]$\lambda\lambda$4959/5007 lines as well as H$\gamma$,
H$\delta$, [NeIII]$\lambda$3869, and [OII]$\lambda$3727
with a redshift of 0.573.
This object is also identified by Rosati et al. (1999) and
in  Stern et al. (2002) as an AGN  
(ID number of 39 by Stern et al. 2002).
Stern et al. (2002) show its optical spectrum with a better
S/N and we do not show our spectrum here.
No.\ 13 is identified with a QSO at $z=1.260$ with  broad
MgII$\lambda$2800, CIII]$\lambda$1909, and CIV1549 emission
lines.
Its spectrum is shown in Figure~\ref{fa2}.
The properties are also summarized in Table~2.
We note that No.\ 5 is identified with a cluster of galaxies
at $z=0.57$ by Vikhlinin et al. (1998) and by Rosati et al. (1999),
and a more detail study of this cluster is presented by
Holden et al. (2001).

\section{A QSO -- SUPERCLUSTER ASSOCIATION AT $z=1.27$?} 
We found three QSOs at $z=1.260-1.286$ in the surveyed
field of $\sim 20^{\prime} \times 20^{\prime}$, which spans
 $\sim 10$ Mpc $\times 10$ Mpc at $z=1.27$
 ($\sim 23$ Mpc at the present epoch).
The redshift difference of $\Delta z =0.026$ corresponds to
a comoving depth of 54 Mpc. 
An expected number of QSOs with X-ray luminosities larger than
$10^{44.5}$ erg s$^{-1}$ (2--10 keV) in this volume is $\sim 0.03$,
if we adopt the X-ray luminosity function in the 2--10 keV band 
at $z=0.8-1.6$ recently obtained by Ueda et al. (2003).
Thus the density of QSOs in this field is very much  high.
Similar overdensities of X-ray sources are found by for example
Gilli et al. (2003) in the \chandra\ deep field south.
They found  density peaks in a region of physical size  7 Mpc
with a depth of $\Delta z < 0.02$, although the X-ray
luminosities seem to be small when compared with our case.
They also point out a correlation between spikes in 
distribution of X-ray sources and those of galaxies
surveyed in the $K$-band.

\begin{figure}
\plotone{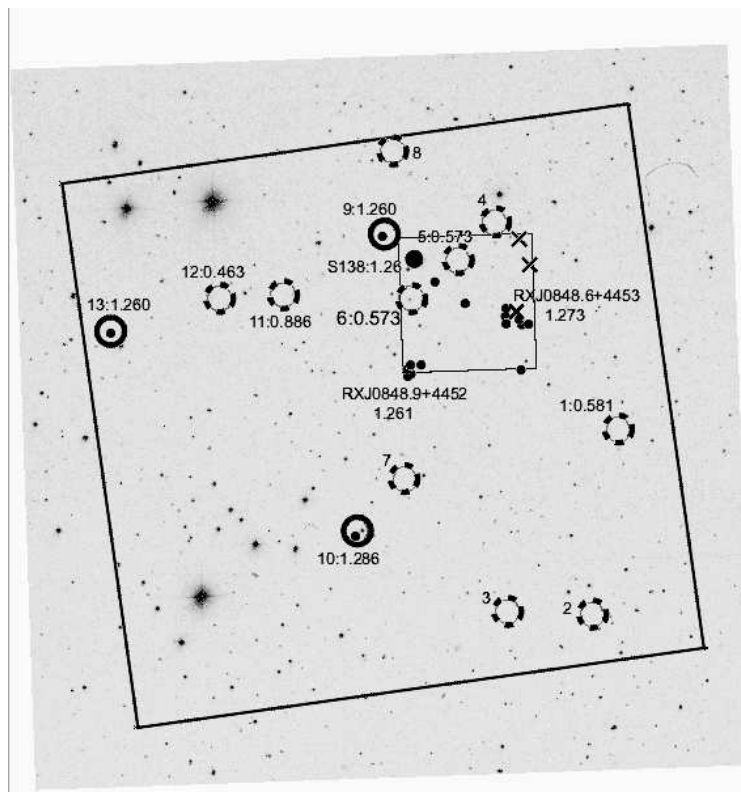}
\caption{
Distribution of X-ray sources and the two rich clusters 
of galaxies (RX J0848.6+4453 at $z=1.27$ and RX J0848.9+4452
at $z=1.26$).
The field of view is $30^{\prime} \times 30^{\prime}$, with
North at the top and East to the left.
The large square shows the field where the \asca\ Lynx survey was 
made. Circles show the X-ray sources detected by \asca\
with their ID numbers and redshifts;
solid circles with a dot (which shows a position of the
optical counterpart) represent QSOs at $z\sim1.3$, while
dotted circles refer to other X-ray sources.
A small square shows the field where the SPICES Deep JK$_{\rm S}$
survey was made (Stern et al. 2002).
Small filled circles refer to spectroscopically confirmed
galaxies in the two clusters (Stanford et al. 1997; Rosati et al. 
1999).
A large filled circle and crosses are an early-type galaxy at
$z=1.26$ (Stern et al. 2002, ID number 138) and red objects
with $R-K_{\rm S} > 5$(Stern et al. 2002), respectively.
\label{fa3}}
\end{figure}

In the Lynx field, the presence of two rich clusters
at $z=1.26$ and 1.27 is known (Stanford et al. 1997;
 Rosati et al. 1999).
The positions (J2000) of these two clusters are RA$=8^{\rm h}
48^{\rm m} 56.^{\rm s}2$ and Dec$=+44^{\circ} 52^{\prime}
0^{\prime\prime}$, and RA$= 8^{\rm h} 48^{\rm m} 34.^{\rm s}2$
and Dec$= 44^{\circ} 53^{\prime} 35^{\prime\prime}$, respectively.
The distribution of the three QSOs and the two clusters (cluster
members) are shown in Figure~\ref{fa3} together with the positions
of other \asca\ sources.
Angular separtions of the three QSOs from these clusters are
$\sim 6^{\prime}-15^{\prime}$,
corresponding to $\sim  3-7.5$ Mpc at the distance and to
comoving separations of $\sim 7 - 17$ Mpc.
Furthermore, seven groups of red galaxies with $z_{\rm ph} = 1-1.35$
are found around the two rich clusters (Nakata et al. 2002).
No.\ 9 is located at an edge of their group 2 (gr2) and No.\ 10
near their group 3 (gr3).
No.\ 13 is out of their field (25$^{\prime} \times 25^{\prime}$).
These groups seem constitute  a supercluster together with the
two rich clusters (Nakata et al. 2002) and
with the QSOs which reside on the outskirts, avoiding the central region
of the supercluster.

Associations of QSOs and clusters were reported by
Tanaka et al. (2001) and by Haines et al. (2001) at $z=1.1-1.2$.
In these studies, QSOs exhibit a  rather clumpy distribution, with a
scale of 20$^{\prime}$--30$^{\prime}$ in a wider $2-4$ degree field.
Clusters are found close to the QSO structure 
and have a scale of 1$^{\prime}$--2$^{\prime}$.
The feature is similar to our case, though 
the luminosities of the QSOs in the Lynx field are slightly smaller.
More recently, a similar trend has been found by
Pentericci et al. (2002) at $z=2.16$; their field contains
two or more QSOs in a $\sim 7^{\prime} \times 7^{\prime}$
field centered on a radio galaxy.
Their field size is smaller than our field and 
their AGNs have slightly smaller X-ray lumiosities than our
three QSOs.
In the lower redshift universe ($z<1$), similar associations
of AGN/QSOs with a cluster have been claimed
(e.g., Cappi et al. 2001; Martini et al. 2002;
Molnar et al. 2002).
The assocication of QSOs and  rich clusters on  large
scales seems to be a rather general feature, suggesting
that  detailed studies could shed light on the link between
star-formation activity and the environmental effect of QSO activity.

\vspace{0.5cm}
This research has made use of data obtained through the High Energy
Astrophysics Science Archive Research Center Online Service, provided
by the NASA-Goddard Space Flight Center. 
Optical follow-up program was supported by grants-in-aid
from the Ministry of Education, Science, Sports  
and Culture of Japan (06640351, 08740171, 09740173)
and from the Sumitomo Foundation.

\clearpage

\begin{deluxetable}{lrrcccc}
\small
\tablenum{1}
\tablecaption{The X-ray Source List\label{tbl-1}} 
\tablehead{
\colhead{No.\tablenotemark{a}} 
&\multicolumn{2}{c}{Position (J2000)\tablenotemark{b}} 
&\multicolumn{3}{c}{Flux ($10^{-13}$ \ergs)\tablenotemark{c}} 
&\colhead{Significance}\\
\colhead{} &\colhead{R.A.} &\colhead{DEC.}
&\colhead{0.7-7 keV} &\colhead{0.7-2 keV} &\colhead{2-7 keV}
&\colhead{0.7-7/0.7-2/2-7 keV}\\
}
\startdata
1*& 132.0404& 44.8206& 0.80$\pm$0.10& 0.22$\pm$0.04& 0.70$\pm$0.12& 8.4/5.9/6.1\nl
2*& 132.0667& 44.6926& 0.36$\pm$0.09& 0.07$\pm$0.03& 0.44$\pm$0.11& 4.3/2.1/4.0\nl
3$^\dag$& 132.1491& 44.6947& 0.34$\pm$0.08& 0.11$\pm$0.03& 0.25$\pm$0.10& 4.3/3.5/2.6\nl
4& 132.1604& 44.9643& 0.37$\pm$0.09& 0.12$\pm$0.03& 0.25$\pm$0.10& 4.3/3.4/2.5\nl
5& 132.1967& 44.9381& 0.34$\pm$0.09& 0.17$\pm$0.04& 0.09$\pm$0.09& 3.8/4.1/1.0\nl
6& 132.2421& 44.9114& 0.64$\pm$0.09& 0.26$\pm$0.04& 0.27$\pm$0.09& 7.5/7.1/3.0\nl
7& 132.2492& 44.7863& 0.31$\pm$0.09& 0.07$\pm$0.03& 0.32$\pm$0.11& 3.6/2.3/2.9\nl
8& 132.2604& 45.0135& 0.59$\pm$0.15& 0.13$\pm$0.06& 0.65$\pm$0.19& 4.0/2.1/3.4\nl
9*& 132.2692& 44.9564& 0.72$\pm$0.10& 0.26$\pm$0.04& 0.45$\pm$0.12& 7.0/5.9/3.9\nl
10*& 132.2967& 44.7502& 0.64$\pm$0.09& 0.24$\pm$0.03& 0.36$\pm$0.10& 7.5/6.8/3.7\nl
11*$^\dag$& 132.3679& 44.9142& 0.43$\pm$0.08& 0.07$\pm$0.03& 0.54$\pm$0.10& 5.6/2.3/5.4\nl
12*& 132.4304& 44.9115& 1.29$\pm$0.10& 0.46$\pm$0.04& 0.83$\pm$0.12& 12.7/10.8/7.2\nl
13& 132.5367& 44.8889& 0.81$\pm$0.14& 0.30$\pm$0.06& 0.50$\pm$0.16& 5.9/5.2/3.1\nl

\tablenotetext{a}{Source number defined in this paper. Sources with * belong
to the hard-band selected sample (above 3.5$\sigma$ in the 2--7 keV band).}
\tablenotetext{b}{The positions from the WGA {\it ROSAT}-PSPC catalog 
(White et al. 1994) except for those flagged with $^\dag$, which 
are detected only with \asca . The \asca\ positional error
is conservatively $0.'5$ in radius.}
\tablenotetext{c}{Flux in each energy band corrected for Galactic
absorption of \nh\ = $2.6\times10^{20}$ cm$^{-2}$.
It is converted from the observed count rate in the same
band assuming a photon index of 1.6.}
\enddata
\end{deluxetable}

\clearpage

\begin{deluxetable}{rccclccccc}
\tabletypesize{\scriptsize}
\tablewidth{0pc}
\tablenum{2}
\tablecaption{Optical Counterparts of the X-ray
Sources\label{tbl-2}} 
\tablehead{
\colhead{No.\tablenotemark{a}} 
&\multicolumn{2}{c}{Position (J2000)\tablenotemark{b}} 
&\colhead{$I_{c}$}
&\colhead{$z$}
&\colhead{Type}
&\colhead{$N_{\rm H}$\tablenotemark{c}}
&\colhead{Photon}
&\colhead{$L_{\rm x}$ (2-10 keV)\tablenotemark{d}}
&\colhead{ID number}\\
\colhead{} &\colhead{R.A.} &\colhead{DEC.}
&\colhead{(mag)} 
&\colhead{}
&\colhead{}
&\colhead{($10^{22}$ cm$^{-2}$)}
&\colhead{index\tablenotemark{c}}
&\colhead{($10^{44}$ erg s$^{-1}$)}
&\colhead{Stern et al. (2002)}\\
}
\startdata
1*  & 8:48:09.9 & $+$44:49:03 & 20.5 & 0.581  & 1 &
 $0.16_{-0.16}^{+0.23}$ & $1.72_{-0.19}^{+0.20}$ & 1.1 & 77 \nl
9*  & 8:49:05.0 & $+$44:57:15 & 19.6 & 1.260  & 1 &
 $0_{-0}^{+0.12}$ & $2.01_{-0.08}^{+0.09}$ & 4.8 & 31 \nl
10* & 8:49:11.4 & $+$44:44:54 & 20.2 & 1.286  & 1 & 
 0 (fixed) & $1.91_{-0.30}^{+0.34}$ & 3.9 & -  \nl
11* & 8:49:27.7 & $+$44:54:58 & 20.2 & 0.886 & 2 & 
 $22_{-4}^{+3}$ & $1.93_{-0.25}^{+0.27}$  & 4.5 & 12 \nl
12* & 8:49:43.3 & $+$44:54:32 & 18.6 & 0.463  & 1 & 
 $0.37_{-0.30}^{+0.32}$ & 1.9 (fixed)  & 0.81 & - \nl
   &          &              &      &        &        &  &  & & \nl
6 & 8:48:58.1 & $+$44:54:35 & 19.0 & 0.573  & 1 &  
  0 (fixed) & $2.26_{-0.35}^{+0.43}$ & 0.45 & 39 \nl
13 & 8:50:09.0 & $+$44:53:16 & 20.9 & 1.260 & 1 & 
  $0.58_{-0.58}^{+2.10}$ & 1.9 (fixed) & 5.1 & - \nl

\enddata
\tablenotetext{a}{Source with an asterisk (*) is a hard band selected
object.}
\tablenotetext{b}{Internal error of the possition is $\sim
0.^{\prime\prime}4$.  Our declination values are systematically 
larger ($\sim 1^{\prime\prime}$) than those of \chandra\ sources
by Stern et al. (2002), presumably due to a differnce of
adopted astrometric catalog.}
\tablenotetext{c}{
$N_{\rm H}$ and photon index for No.\ 1, No.\ 9, and
No.\ 11 are derived using \chandra\ archive data (see the text).  For
other sources, either $N_{\rm H}$ or photon index is derived with the
procedure adopted by Ueda et al. (2003) from the two-band hardness
ratio of the SIS data (i.e., a value of either $N_{\rm H}$ or photon
 index is assumed). Errors are 1$\sigma$ for a single parameter.
}
\tablenotetext{d}{Intrinsic rest-frame luminosity after correcting
for $N_{\rm H}$, calculated from the SIS 2--7 keV count rate with
the best-fit spectral parameters.}

\end{deluxetable}

\end{document}